\documentclass[letterpaper,english,aps, prl, reprint, superscriptaddress]{revtex4-1}

\usepackage[T1]{fontenc}
\usepackage[latin9]{inputenc}
\usepackage{amsmath}
\usepackage{amssymb}
\usepackage{bbold}
\usepackage{graphicx}

\makeatletter

\pdfpageheight\paperheight
\pdfpagewidth\paperwidth

\PassOptionsToPackage{position=top}{subfig}
\usepackage{hyperref}
\hypersetup{
	colorlinks = true,
	allcolors = blue
}

\usepackage{times}

\makeatother

\usepackage{babel}
\begin{document}

\title{Exact complex mobility edges and flagellate spectra for non-Hermitian quasicrystals with exponential hoppings}
%\title{Exact non-Hermitian mobility edges and flagellate spectra in quasicrystals with exponential hoppings}
%\title{Flagellate spectra in a class of non-Hermitian quasiperiodic lattices with exponential hoppings}
%\title{Flagellate spectra in a class of non-Hermitian quasiperiodic lattices with exponential short-range hoppings}
%\title{Analytical non-Hermitian mobility edges in a class of quasiperiodic lattices with long-range hopping}
%\title{Analytical non-Hermitian mobility edges in a family of quasiperiodic lattices with long-range hopping}
%\title{Analytical non-Hermitian mobility edge loop in a family of quasiperiodic lattices with long-range hopping}

\author{Li Wang}
\email{liwangiphy@sxu.edu.cn}
\affiliation{Institute of Theoretical Physics, State Key Laboratory of Quantum Optics and Quantum Optics Devices, Collaborative Innovation Center of Extreme Optics, Shanxi University, Taiyuan 030006, P. R. China}

\author{Jiaqi Liu}
\affiliation{Institute of Theoretical Physics, State Key Laboratory of Quantum Optics and Quantum Optics Devices, Collaborative Innovation Center of Extreme Optics, Shanxi University, Taiyuan 030006, P. R. China}

\author{Zhenbo Wang}
\affiliation{Institute of Theoretical Physics, State Key Laboratory of Quantum Optics and Quantum Optics Devices, Collaborative Innovation Center of Extreme Optics, Shanxi University, Taiyuan 030006, P. R. China}

\author{Shu Chen}
\email{schen@iphy.ac.cn }
%\thanks{schen@iphy.ac.cn }
\affiliation{Beijing National Laboratory for Condensed Matter Physics, Institute
of Physics, Chinese Academy of Sciences, Beijing 100190, China}
\affiliation{School of Physical Sciences, University of Chinese Academy of Sciences,
Beijing 100049, China }

\date{\today}
%\date{}

\begin{abstract}
We propose a class of general non-Hermitian quasiperiodic lattice models with exponential hoppings and analytically determine the genuine complex mobility edges by solving its dual counterpart exactly utilizing Avila's global theory. Our analytical formula unveils that the complex mobility edges usually form a loop structure in the complex energy plane. By shifting the eigenenergy a constant $t$, the complex mobility edges of the family of models with different hopping parameter $t$ can be described by a unified formula, formally independent of $t$. By scanning the hopping parameter, we demonstrate the existence of a type of intriguing flagellate-like spectra in complex energy plane, in which the localized states and extended states are well separated by the complex mobility edges.
Our result provides a firm ground for understanding the complex mobility edges in non-Hermitian quasiperiodic lattices.
\end{abstract}

\maketitle

\textcolor{blue}{\em Introduction.}--
It is well-known that metal-insulator phase transitions are of paramount significance in the field of condensed matter physics, underlying which, many fundamental mechanisms have been discovered. Among these, disorder-induced localization undoubtedly holds a non-negligible position, which was first identified by Anderson in his seminal work~\cite{Anderson1958pr}, and now has evolved into an enduring and ever-renewing research topic. In three-dimensional (3D) systems subjected to truly random disorder, energy-dependent localization may occur~\cite{Mott1967,Mott1987jpc}. However, for low-dimensional systems (1D and 2D), disorder of arbitrarily small strength can render all eigenstates of the system into localized states, thus preventing the emergence of finite disorder-induced transitions~\cite{Abrahams1979prl,RevModPhys.57.287,RevModPhys.80.1355}.
Upon this, the Aubry-Andr\'{e}-Harper (AAH) model~\cite{AA1980,Harper_1955} featuring quasi-periodicity was proposed, which is proven to be able to accommodate extended-localized transitions even in 1D at a finite strength.
This pivotal finding has deeply stimulated the research on localization transitions and the fundamental mobility edge physics in low-dimensional systems~\cite{Thouless1972,Thouless,Kohmoto,Kohmoto2008,Cai2013,Roati,Lahini,Bloch,prb10817,wang2024ME,
li2024ring}.
Various extensions~\cite{xiexc1988prl,xieprb415544,Hiramoto,HanJ,Biddle10prl,Ganeshan2015prl,DengX,
SciPostPhys.13.3.046,mprb23,vu224206,m23critical,SciPostPhys.12.1.027,XuZ,YaoH,LiX} of the AAH model have been put forward, some focusing on the short-range~\cite{Roy21prl} or long-range hopping~\cite{Biddle10prl, Biddle11prb, Santos19prl, prb103075124}, while others are committed to constructing delicate quasiperiodic potentials~\cite{xiexc1988prl,xieprb415544,Ganeshan2015prl,Lixp16prb,Sarma17prb,Lix20prb,mosaic-exp,
prl125196604,prb105L220201,prb96174207,prb10817}, aiming at the emergence of mobility edge. Due to the discovery of duality relations~\cite{AA1980,Ganeshan2015prl,prb10817} and Avila's global theory~\cite{avila,zhouqiwang2023}, quasicystals has gained a remarkable advantage, which grants researchers the opportunity to explore rigorous results~\cite{Ganeshan2015prl,Biddle10prl,Biddle11prb,wyc20prl,LXJprl131.176401}.

In recent years, non-Hermitian physics has entered a booming period and attracted intensive studies, in which some novel physics and exotic phenomena have emerged~\cite{bender98,RPP70947,NP1411,AP69249,NP131117,prl121086803,prl121.026808,JPC2035043,PRX8031079,
PRX9041015,PRB99201103,PRL123066404,NP16761,PRL125126402,
PRL124086801,CPB30020506,prl128120401,PRL124056802,PRL125226402,PRL127116801,NC115491,RMP93015005,PRX13021007}.
The interplay of non-Hermiticity and disorder brings a new perspective for the localization phenomena, leading to increasing studies of extended-localized transitions within non-Hermitian systems \cite{HN,Satija,JiangH,Longhi2019,LiuYX2021a,Liuyx2021,PhysRevB.101.174205,HuHui,
PhysRevB.103.214202,Chen21prb,ZengQB,PhysRevB.100.125157,PhysRevB.106.024202,CaiXM2021,
Longhi2021,XuZ2021,PhysRevB.105.054204,PhysRevB.105.054201,PhysRevB.106.144208,
ZhouLW,XiaX2022,Datta,Gandhi,Mishra,XueP,Datta2024}.
From the perspective of the underlying mathematical foundations, the topic has expanded from Hermitian random matrices to non-Hermitian random matrices. While Hermitian random matrices contain $10$ symmetry classes according to Altland-Zirnbauer classification, non-Hermitian random matrices possess $38$ classes based on Bernard-LeClair classification~\cite{S207,PRX9041015,PRB99235112,PRB100144106}. Compared to Hermitian counterpart, the spectral statistics of non-Hermitian disordered systems exhibit distinct features~\cite{PRL802897,J42265204,PRL83484,PRL813367}.
Previous studies~\cite{Chen21prb,PhysRevB.101.174205,PhysRevB.103.214202,Mishra,Longhi2019,HuHui,Gandhi,
Mishra,XueP,LiuYX2021a,Liuyx2021,Datta,ZhouLW,XiaX2022} have mainly focused on parity-time ($\mathcal{PT}$) symmetric~\cite{bender98} quasiperiodic systems, for which an appealing and sound correspondence between real-complex transition in eigenenergy and energy-dependent extended-localized transition exists.
Thus, although the Hamiltonian is non-Hermitian, mobility edge is still described in real domain except few numerical attempts~\cite{PhysRevB.106.144208,ZhouLW}.
Very recently, analytical genuine \emph{complex mobility edges} for general non-Hermitian 1D quasicrystals have been addressed in Ref.~\cite{wang2024ME,li2024ring}, in which non-Hermitian quasiperiodic models with only nearest-neighbor hopping are investigated, and the quasiperiodic potentials are delicately designed.

\begin{figure*}[tp]
\includegraphics[width=18cm]{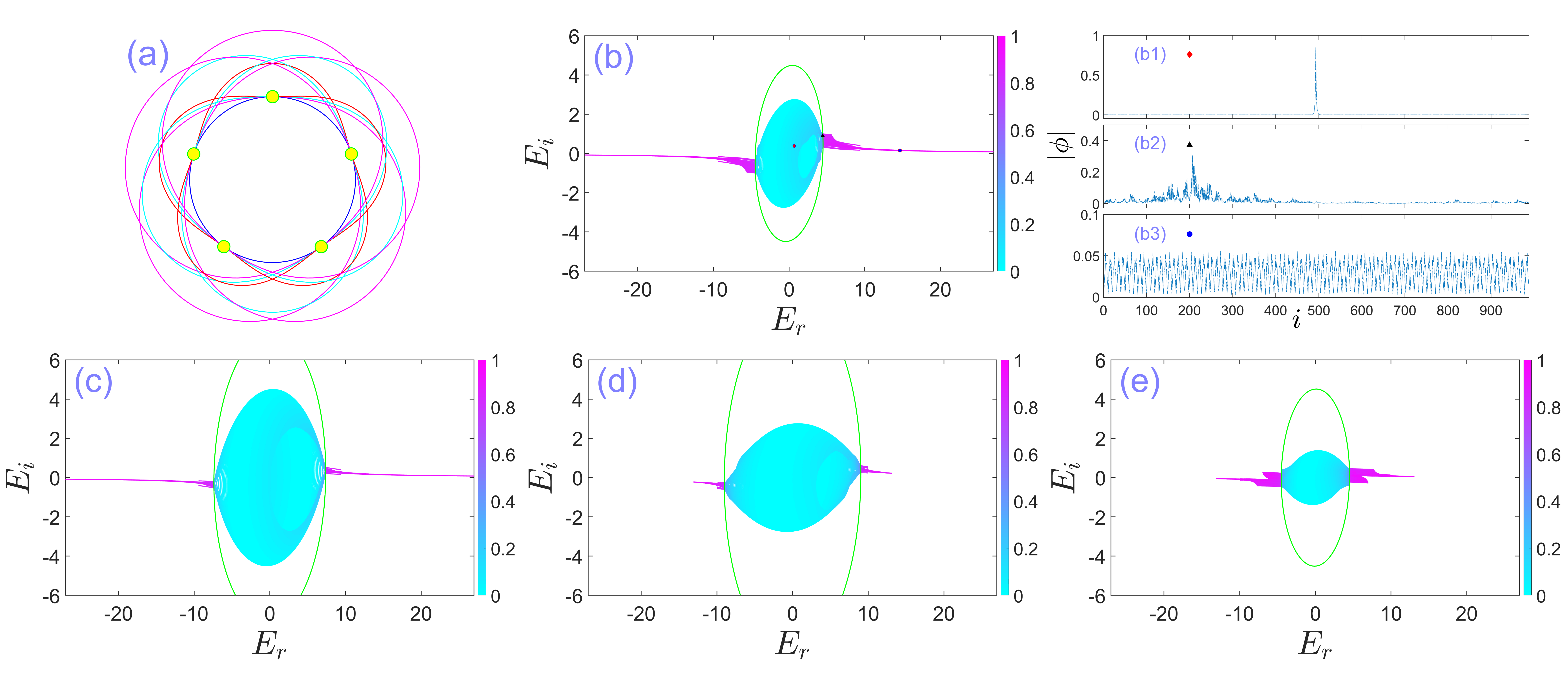}
\caption{(a) Sketch of the class of 1D non-Hermitian quasiperiodic lattice models with exponential hoppings under periodic boundary conditions (PBCs).
Each circular area in yellow color denotes a lattice site. Connecting lines in different colors represent exponential hopping processes at different distances $|j-j'|$.
(b-e) Flagellate spectra of the class of non-Hermitian quasiperiodic lattices with exponential hoppings. Models with different reference hopping amplitudes $t$s share common non-Hermitian mobility edges (NHMEs). Their common NHMEs are given precisely by Eq.~(\ref{LRAANHME}) and are vividly represented by the solid green line.The color of each energy point denotes the fractal dimension (FD) of the corresponding eigenstate. (b) $p=0.5$, $h=1$, $V=2$. (c) $p=0.5$, $h=1.5$, $V=2$. (d) $p=1.2$, $h=1$, $V=2$. (e) $p=1.2$, $h=1$, $V=1$. Other common parameters are, $L=987$, $\chi=\pi/4$.
(b1-b3) Representative spatial distributions of eigenstates in different parts of the non-Hermitian flagellate spectra in (b).
(b1) Localized state indicated by a red diamond in (b).
(b2) Critical state near NHMEs corresponding to eigenenergy denoted by a black triangle in (b).
(b3) Extended state corresponding to eigenenergy marked by a blue dot in (b).}
\label{Fig01}
\end{figure*}
% $F\!D$

In this work, we propose a class of non-Hermitian quasiperiodic models with exponential hoppings and solve it exactly to obtain a compact analytical formula for complex mobility edges, generalizing the topic to models with non-nearest-neighbour hoppings.
Notably, by combining an extra on-site (zero-distance hopping) term into the exponential hopping terms in Eq.~(\ref{exphop}), which is equivalent to making a shift of the eigenvalues, the genuine complex mobility edges do not change with the variation of hopping parameter $t$.
When $t$ scans the parameter region, a type of exotic flagellate-like spectra appear in the complex energy plane, which share common complex mobility edges.
With a typical example, we demonstrate that the localized and extended states distribute on the body  and the flagellum of the \emph{flagellate}, respectively, separated by complex mobility edges.
In general, a model with long-range hoppings is difficult to deal with, thus we derive the exact complex mobility edges by exactly solving its dual model, which is a non-Hermitian extension of the Ganeshan-Pixley-Das Sarma (GPD) model by introducing in both non-reciprocal hopping and complex potential.
The Lyapunov exponent of the dual model can be analytically obtained by implementing Avila's global theory~\cite{avila,zhouqiwang2023}, which allow us to arrive at an analytical formula of the complex mobility edges for the dual model. Thus, the compact analytical formula for the exact complex mobility edges of the class of models described by Eq.(\ref{exphop}) is obtained according to the parameter mappings between the model and its dual.

\textcolor{blue}{\em Model and structure of mobility edges.}--We study localization transitions in a class of general non-Hermitian quasiperiodic lattices with exponential hopping terms, which can be briefly described by the following eigenvalue equation,
\begin{equation}
\sum_{j'} t e^{-p|j-j'|} \phi_{j'}+ V e^{i \chi} \cos (2\pi \alpha j +i h)\phi_j =E \phi_j,
\label{exphop}
\end{equation}
where $j$ and $j'$ are indices of lattice sites, $t$ is the reference hopping amplitude and $p>0$ denotes the decay rate of the exponential hoppings.
Here, the summation of $j'$ includes the term $j'=j$. Making a shift of $E$ by $E+t$ can exclude the term of $j'=j$.
An intuitive and visual schematic representation of this model is provided in Fig.~\ref{Fig01}a.
Depending on the value of decay rate $p$, the model can smoothly interpolate between truly long-range and short-range.
The parameter $\alpha$ is an irrational number which is responsible for the quasiperiodicity of the lattice potential. To be concrete and without loss of generality, we set $\alpha=(\sqrt{5}-1)/2$ in this work.
A key feature of the model is the seamless integration of two different kinds of non-Hermiticity into one unified framework, which are dictated by $\chi$ and $h$, respectively. The distinct nature of these two non-Hermitian terms will be further illuminated in the dual model, which will be elaborated upon in subsequent sections.
It's not hard to see that a non-Hermitian potential of this particular form typically does not adhere to parity-time ($\mathcal{PT}$) symmetry, which has been a crucial aspect of earlier studies on non-Hermitian localization transitions~\cite{Chen21prb,PhysRevB.101.174205,PhysRevB.103.214202,Mishra,Longhi2019,HuHui,Gandhi,Mishra,
XueP,LiuYX2021a,Liuyx2021,Datta,ZhouLW,XiaX2022}. For the Hermitian limit with $\chi=0$ and $h=0$, the model becomes the Biddle-Das Sarma (BD) model~\cite{Biddle10prl,Biddle11prb} after a shift of $E$ by $E+t$, where an analytical formula for the mobility edge is obtained through implementing a self-dual transformation. The model with $\chi=0$ and non-zero $h$ is $\mathcal{PT}$-symmetric and has a real mobility edge~\cite{Chen21prb}.
For the general case with $\chi \neq 0$ and $h \neq 0$, %and without the constraint of $j\neq j'$,
no analytical result is obtained yet.

In this study, we analytically derived the precise non-Hermitian mobility edges for the general cases of the model outlined in Eq.(\ref{exphop}). These complex mobility edges, which divide localized and extended states in the complex plane, can be analytically expressed as,
\begin{align}
\frac{( E_r\cos\chi + E_i\sin\chi )^2}{\cosh^2 (p+|h|)}+\frac{ ( E_r\sin\chi - E_i\cos\chi )^2 }{\sinh^2 (p+|h|)}=V^2,
\label{LRAANHME}
\end{align}
where $E_r$ and $E_i$ are the real and imaginary parts of $E$, respectively.
Eq.~(\ref{LRAANHME}) demonstrates that under normal circumstances, the mobility edges take complex values.
When $h=0$, $\chi=0$, and discarding the zero-distance hopping term $t\phi_j$, the model Eq.~(\ref{exphop}) reduces to its Hermitian limit with real $E$ and Eq.~(\ref{LRAANHME}) is simplified to
$E_r=\pm V\cosh(p)$, consistent with the results of Refs.~\cite{Biddle10prl,Biddle11prb} by substituting $E_r=E+t$.
Interestingly, observing Eq.~(\ref{LRAANHME}), one can find that the exact complex mobility edges form a closed elliptic loop, featuring localized eigenstates within its interior and extended eigenstates outside.
Specifically, the ellipse, with its center located at the origin of the complex plane, has a major axis length of $|V|\cosh(p+|h|)$, a minor axis length of $|V|\sinh (p+|h|)$. Thus, $|V|$, $|h|$, and the decay rate $p$ together determine the size of the ellipse's area, which delineates the possible distribution range for eigenenergies of the localized states. And the non-Hermitian parameter $\chi$ determines the orientation of the semi-major axis.

The compact analytical formula Eq.~(\ref{LRAANHME}) for the complex mobility edges of the general non-Hermitian model Eq.~(\ref{exphop}) can be further verified by straightforward numerical simulations.
We exploit fractal dimension (FD) as a criterion to distinguish between localized and extended eigenstates. For an arbitrary normalized eigenstate $\phi$, the fractal dimension is defined as $F\!D=-\lim_{L\rightarrow \infty} \ln(\sum_j |\phi_j|^4)/\ln L$. It is clear that for localized states, $F\!D\!\rightarrow\!0$ and for extended states, $F\!D\!\rightarrow\!1$.
It is worth noting that, according to Eq.~(\ref{LRAANHME}), the non-Hermitian mobility edges of the model are independent of the reference hopping amplitude $t$. This is an appealing feature in that models with different reference hopping amplitudes $t$s share same non-Hermitian mobility edges.
As a concrete example, we numerically calculate the spectra for the non-Hermitian quasiperiodic lattices under periodic boundary condition (PBC) with $L=987$, $p=0.5$, $h=1$, $V=2$ and $\chi=\pi/4$, as depicted in Fig.~\ref{Fig01}(b). As all the energy spectra of models with varying reference hopping amplitude $t$, ranging from $-7$ to $7$, are plotted on the same complex plane, a novel and interesting flagellate-like energy spectrum structure emerges.
Remarkably, for the \emph{flagellate}, its flagellum, as its locomotor organ, corresponds to extended states with high mobility, while the body part inside the ellipse corresponds precisely to localized states with low mobility.

It can be inferred that when the decay rate $p$ increases, the effective range of exponential hoppings will decrease rapidly, and the proportion of localized eigenstates in the flagellate spectra will increase. This is vividly demonstrated in Fig.~\ref{Fig01}d. Comparing to Fig.~\ref{Fig01}b, the decay rate $p$ increased from $0.5$ to $1.2$, while other parameters are the same. At the same time, the complex mobility edge loop has been significantly enlarged. Alternatively, increasing $h$ can also elevate the ratio of localized eigenstates in the flagellate spectra, as shown in Fig.~\ref{Fig01}c. The portion inside the complex mobility edge loop is notably expanded.
This is due to the fact that increasing $h$ is equivalent to enhancing the strength of the quasiperiodic potential. Also, the green complex mobility edge loop has expanded considerably.
In contrast, if the strength $V$ of the on-site quasiperiodic potential is decreased, the ratio of the extended eigenstates is anticipated to rise, which is clearly shown in Fig.~\ref{Fig01}e. The model parameters in Fig.~\ref{Fig01}e are the same as those in Fig.~\ref{Fig01}d except $V$ has decreased from $2$ to $1$. Obviously, the portion of extended states is increased compared to Fig.~\ref{Fig01}d. Meanwhile, the size of the complex mobility edge loop shrinks.

Finally, to obtain some intuitive visualizations, we present eigenstates in Fig.~\ref{Fig01}(b1-b3) for $L=987$, $V=2$, $p=0.5$, $\chi=\pi/4$, $h=1$, and $t=5$ at three different energy eigenvalues which lie inside, near, and outside the complex mobility edge loop.
We see that the wave function in Fig.~\ref{Fig01}(b1) is localized for the energy eigenvalue marked by a red diamond in the flagellate spectra displayed in Fig.~\ref{Fig01}b, while the wave function in Fig.~\ref{Fig01}(b3) is extended for the energy eigenvalue marked by a blue point.
The wave function in Fig.~\ref{Fig01}(b2) is critical with multifractal structure for the energy eigenvalue marked by a black triangle.

\textcolor{blue}{\em Derivation of exact mobility edges.}
--Now we present the detailed analytical derivation of the exact non-Hermitian mobility edges Eq.~(\ref{LRAANHME}) for the generic non-Hermitian quasiperiodic models  with exponential hoppings.
Rather than directly tackling the original model presented in Eq.~(\ref{exphop}), we will instead address the corresponding dual model.
Implementing the following transformation
\begin{equation}\label{transform}
\phi_j=\sum_{n}^{}e^{i 2\pi \alpha jn} u_n
\end{equation}
on Eq.~(\ref{exphop}), we arrive at a general non-Hermitian extension of the well known Ganeshan-Pixley-Das Sarma (GPD)  model \cite{Ganeshan2015prl, prb10817,Cai_2023,wang2024ME,li2024ring},
\begin{equation}\label{lamdelta}
e^{h}u_{n-1}+e^{-h}u_{n+1}+\frac{\lambda e^{-i\chi}\cos(2\pi\alpha n)}{1-b \cos(2\pi\alpha n)} u_n=  \epsilon u_n,
\end{equation}
in which,
\begin{align}
b&=\mathrm{sech}(p),\label{bb}\\
\lambda&=\frac{2t \tanh(p)}{V\cosh(p)}, \label{lam} \\
\epsilon&=\frac{2[E-t\tanh(p)]}{V e^{i\chi}}. \label{epsl}
\end{align}
According to the transformation Eq.~(\ref{transform}), it is clear that the model in Eq.~(\ref{lamdelta}) is dual to the one described by Eq.~(\ref{exphop}).
Notably, the non-Hermiticity of model Eq.~(\ref{lamdelta}) is introduced in  through two distinct ways: one is through nonreciprocal hopping, controlled by paramter $h$, and the other is via gain and loss, represented by complex on-site potential, dictated by parameter $\chi$. This model perfectly amalgamates both non-Hermitian mechanisms within a unified framework.

The reciprocal limit of Eq.~(\ref{lamdelta}) with $h=0$ has been analytically tackled in Ref.~\cite{wang2024ME} through computing Lyapunov exponent $\gamma$ exactly utilizing Avilia's global theory of one-frequency analytical $S\!L(2,\mathbb{C})$ cocycle\cite{avila,zhouqiwang2023}.
Specifically, in the context of this work, the Lyapunov exponent is written as
\begin{align}
\gamma(\epsilon)=\mathrm{max}\{\ln F , 0\}, \label{LEh0}
\end{align}
where
\begin{align}
F=\mathrm{max}\left \{ \left | \frac{-(b\epsilon+\lambda e^{-i\chi}) \pm \sqrt{(b\epsilon+\lambda e^{-i\chi})^2-4 b^2}}{2(1+\sqrt{1-b^2})}\right|\right \}.
\end{align}

Based on the analytical formula for the non-Hermitian mobility edges given in Ref.~\cite{wang2024ME} and replacing the corresponding quantities according to the mapping given by Eq.~(\ref{bb})-(\ref{epsl}), one can straightforwardly obtain the exact non-Hermitian mobility edges of the model in Eq.~(\ref{exphop}) under the condition of $h=0$, which reads,
\begin{align}
\frac{( E_r\cos\chi + E_i\sin\chi )^2}{\cosh^2 (p)}+\frac{ ( E_r\sin\chi - E_i\cos\chi )^2 }{\sinh^2 (p)}=V^2.
\label{NHMEh0}
\end{align}

However, for the most general case of Eq.~(\ref{lamdelta}) with $h\neq 0$, one has to further generalize the derivations in Ref.~\cite{wang2024ME} to accommodate the non-reciprocal hoppings.
Under open boundary conditions (OBCs), the non-reciprocal hoppings generally may lead to non-Hermitian skin effect~\cite{prl121086803,prl121.026808,JiangH}. %prl123.066404,prl124.056802
Apparently, the non-Hermtian Hamiltonian $H$ corresponding to Eq.~(\ref{lamdelta}) reads,
\begin{align}
H=\sum_{n} \left ( e^{h}|n+1 \rangle \langle n|+e^{-h}|n \rangle \langle n+1| \right ) \nonumber \\
+\sum_{n}\frac{\lambda e^{-i\chi}\cos(2\pi\alpha n)}{1-b \cos(2\pi\alpha n)} |n \rangle \langle n|, \label{Hld}
\end{align}
which under OBC can be transformed to a Hermitian Hamiltonian $\bar{H}$ through similar transformation $\bar{H}=SHS^{-1}$. $S$ is the similarity matrix with exponentially decaying diagonal elements, i.e. $S=(e^{-h}, e^{-2h},..., e^{-Lh})$. Simple calculations reveal that $\bar{H}=H(h=0)$, which implicates that the similar transformation reduces the non-reciprocal Hamiltonian $H$ to a reciprocal one without skin modes. The eigenstates of $H$ and $\bar{H}$ are related by $|u \rangle =S^{-1}|\bar{u} \rangle$. Thus, the similarity transformation matrix $S^{-1}$ plays the role of exponentially accumulating probability towards boundary. Generally, it can successfully transform an extended eigenstate into a skin mode but fails to affect the localized state since the probability of a localized state is highly confined \cite{LiuYX2021a}.

Within the scenario of Anderson localization, the modulus of an eigenstate of $\bar{H}$ can generally be represented as $|\bar{u}_n|\propto e^{-\gamma |n-n_0|}$, where $n_0$ corresponds to the center of the probability distribution, and $\gamma$ denotes the Lyapunov exponent (LE) which is the inverse of localization length. Thus, if $\gamma \rightarrow 0$ as $L \rightarrow \infty$, the state $|\bar{u} \rangle$ is extended, otherwise the state is localized. Correspondingly, the eigenstates of $H$ have the form of $|u_n|\propto e^{hn-\gamma|n-n_0|}$, from which one can infer that when $h>\gamma$ ($h<-\gamma$), delocalization occurs on the right (left) side of $n_0$ and skin modes may further emerge on the same side. Accordingly, under open boundary condition the transition point between localized states and skin states is given by $\gamma=|h|$. Since localized state is insensitive to boundary condition. Thus, under periodic boundary condition, the transition point separating localized states and extended states is also $\gamma=|h|$. Substituting Eq.~(\ref{LEh0}) into $\gamma=|h|$, we thus obtain the exact non-Hermitian mobility edges for the general non-Hermitian quasiperiodic model in Eq.~(\ref{lamdelta}), which reads,
\begin{align}
\frac{(b \epsilon_r+\lambda \cos\chi)^2}{(\cosh|h|+A \sinh|h|)^2} +
\frac{(b \epsilon_i-\lambda \sin\chi)^2}{(\sinh|h|+A \cosh|h|)^2} =4, \label{lamNHMEh}
\end{align}
where $A=\sqrt{1-b^2}$. For the $h=0$ limit, Eq.~(\ref{lamNHMEh}) reduced to the result obtained in Ref.~\cite{wang2024ME}.
Replacing $\epsilon$, $\lambda$, and $b$ with $E$, $V$ and $p$, one finally obtains the exact complex mobility edges demonstrated in Eq.~(\ref{LRAANHME}) for the the general non-Hermitian quasiperiodic models with exponential hoppings.

\begin{figure}[bp]
\includegraphics[width=8.7cm]{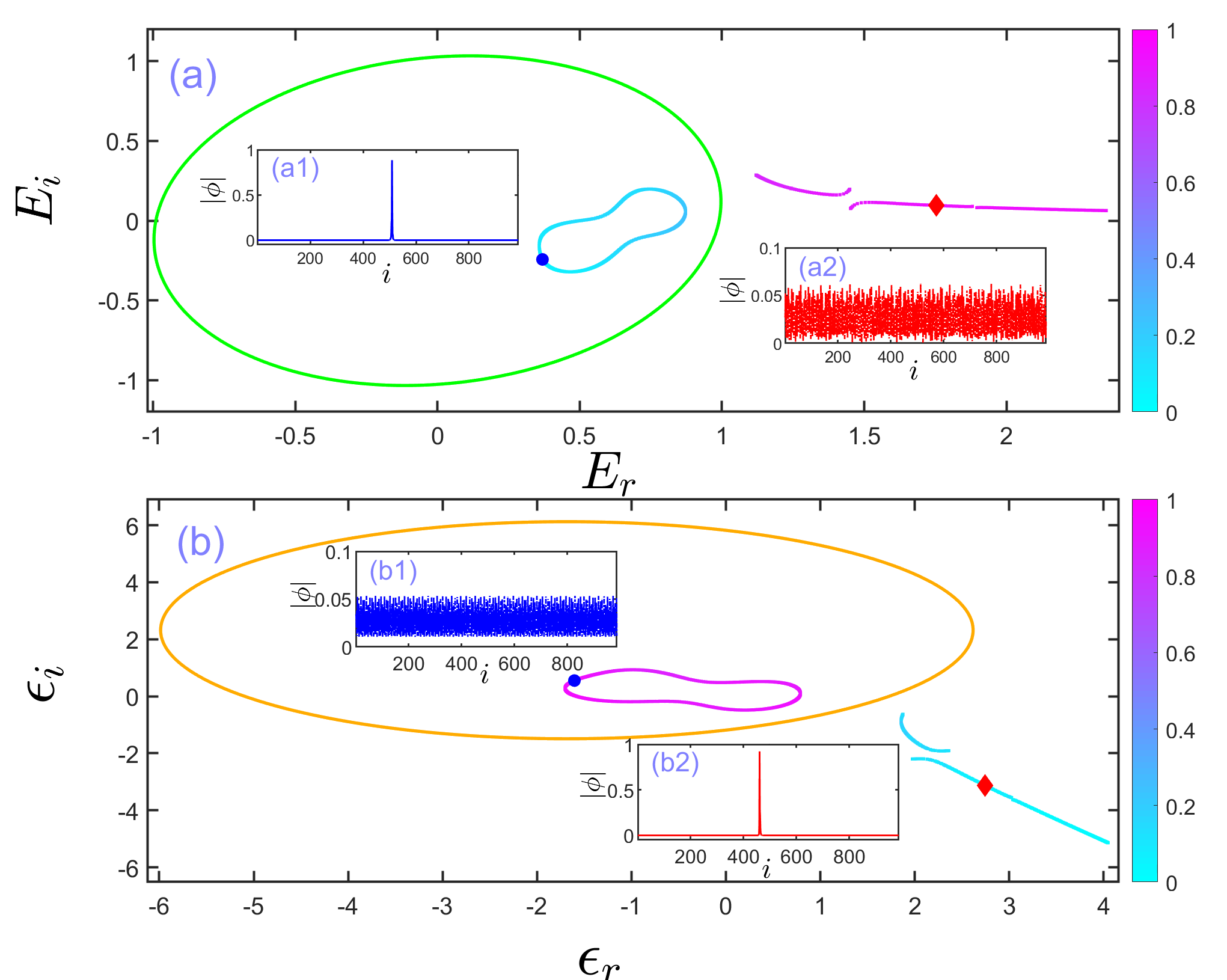}
\caption{\label{NHMEcompare}
Typical spectra and the exact complex mobility edges of the non-Hermitian quasiperiodic lattice model with exponential hoppings and its non-reciprocal dual counterpart.
(a) An example spectrum of the model described by Eq.~(\ref{exphop}) with $V=0.5$, $p=0.9$ and $t=1$. The color of each energy point denotes the fraction dimension of the corresponding eigenvector to visualize its localization property.
The green ellipse is plotted according to the exact complex mobility edge formula Eq.~(\ref{LRAANHME}).
(b) An example spectrum of the non-Hermitian dual model described by Eq.~(\ref{lamdelta}) with $\lambda=1.9993$ and $b=0.69779$. The color of each energy point denotes the fraction dimension of the eigenvector.
The ellipse in orange is given by the exact complex mobility edge formula Eq.~(\ref{lamNHMEh}).
Each inset shows the spatial distribution of the eigenvector corresponding to the marked eigenenergy nearby.
Other parameters in common are $L=987$, $\chi=0.3\pi$ and $h=0.5$.
}\label{Fig04}
\end{figure}

%\textcolor{blue}{\em NHMEs of the dual model.}--
\textcolor{blue}{\em Discussion on the model and its dual counterpart.}--
The model in Eq.~(\ref{exphop}) and its dual model are deeply related as they can transform to each other through the transformation represented by Eq.~(\ref{transform}). Their non-Hermitian mobility edges are both analytical exact and are closely connected.
In Fig.~\ref{Fig04}, we numerically show the representative spectra for both the model with exponential hoppings (Fig.~\ref{Fig04}a) and its dual counterpart (Fig.~\ref{Fig04}b) on finite quasiperiodic lattices with $L=987$ under periodic boundary conditions. Model parameters are $V=0.5$, $p=0.9$, $\chi=0.3\pi$, $h=0.5$, $t=1$ in Fig.~\ref{Fig04}a, and are $\lambda=1.9993$, $b=0.69779$, $\chi=0.3\pi$, $h=0.5$ in Fig.~\ref{Fig04}b.
As can be seen from Fig.~\ref{Fig04}a, the energy spectrum structure of the general non-Hermitian model with exponential hoppings can be divided into two categories, one is the circular structure, and the other is the dendritic structure.
These two kinds of spectrum structures are well separated by the exact non-Hermitian mobility edge loop in green given analytically by Eq.~(\ref{LRAANHME}).
And also, the spectrum of the dual non-reciprocal non-Hermitian model in Fig.~\ref{Fig04}b possesses similar structures. The complex mobility edge loop in orange is plotted according to Eq.~(\ref{lamNHMEh}).
Apparently, the structure of the energy spectrum in Fig.~\ref{Fig04}a and the one in Fig.~\ref{Fig04}b is topologically equivalent. This is sensible in that the eigenvalues of the model and its dual are connected by a mathematical mapping shown in Eq.~(\ref{epsl}) and the mapping is topology-preserving.
It is worth noting that as long as $h$ is nonzero, the energy spectrum within the non-Hermitian mobility edge loop will exhibit closed circular structures.
On the other hand, the model and its dual counterpart are distinct. The energy spectrum with the same kind of structure corresponds to eigenstates with different localization properties in the two models.
For example, the same circular energy spectra in the Fig.~\ref{Fig04}a and Fig.~\ref{Fig04}b exhibit contrasting colors, where the colors represent the fractional dimensions of the corresponding eigenstates. Furthermore, in the insets, we provide specific examples of the spatial distribution of eigenstates corresponding to the eigenenergies in different regions of the energy spectra.

\textcolor{blue}{\em Summary.}--
In summary, we have studied a class of general non-Hermitian quasiperiodic lattice models with exponential hoppings and derived an exact analytical formula for the genuine complex mobility edges. By making a shift of the eigenvalues, the complex mobility edges of the models with different hopping parameter $t$ are formally $t$-independent, and thus a family of models with various $t$ shares the same complex mobility edges. Our numerical results demonstrate that the spectra of the family of models exhibit intriguing flagellate-like pattern in complex energy plane, in which the localized states and extended states are well separated by the common exact complex mobility edges. Remarkably, for the \emph{flagellate}, its flagellum, as its locomotor organ, corresponds to extended eigenstates with high mobility, while the body part inside the complex mobility edge loop corresponds precisely to localized eigenstates with low mobility.
Finally, a comparison is made between the model and its dual counterpart.
This study will further facilitate the extension of the fundamental concept of mobility edge to non-Hermitian physics. And also, these findings clearly reveal the high diversity and complexity of the spectrum structures of non-Hermitian quasicrystals.

\textcolor{blue}{\em Acknowledgments}--
%\begin{acknowledgments}
L.W. is supported by the Fundamental Research Program of Shanxi Province, China (Grant No. 202203021211315), Research Project Supported by Shanxi Scholarship Council of China (Grant No. 2024-004), the National Natural Science Foundation of China (Grant Nos. 11404199, 12147215) and the Fundamental Research Program of Shanxi Province, China (Grant Nos. 1331KSC and 2015021012). S. C. is supported by  by National Key Research and Development Program of China (Grant No. 2023YFA1406704), the NSFC under Grants No. 12174436 and
No. T2121001 and the Strategic Priority Research Program of Chinese Academy of Sciences under Grant No. XDB33000000.
%\end{acknowledgments}

%\bibliography{MyBib}


\end{document}